\documentclass[english,twocolumn,aps,prb,nofootinbib]{revtex4-1}
\usepackage[utf8]{inputenc}
\usepackage[T1]{fontenc}
\usepackage{babel}
\usepackage{amsmath}
\usepackage{graphicx}
\usepackage{fancyhdr}
\usepackage{braket}
\usepackage{xcolor}

\begin{document}

\title{
A braiding quantum circuit based on the $4\pi$ Josephson effect}

\author{John P. T. Stenger, Michael Hatridge, Sergey M. Frolov, David Pekker}
\affiliation{Department of Physics and Astronomy, University of Pittsburgh, Pittsburgh, PA 15260, USA}

\date{\today}

\begin{abstract}
We propose a topological qubit in which braiding and readout are mediated by the $4\pi$ Majorana-Josephson effect. 
The braidonium device consists of three Majorana nanowires that come together to make a tri-junction.  In order to control the superconducting phase differences at the tri-junction, the nanowires are enclosed in a ring made of a conventional superconductor.  In order to perform initialization/readout, one of the nanowires is coupled to a fluxonium qubit through a topological Josephson junction.  We analyze how flux-based control and readout protocols can be used to demonstrate braiding and qubit operation for realistic materials and circuit parameters.

\end{abstract}

\maketitle

\section{Introduction}
Topological quantum computation promises a path towards robust quantum information storage and manipulation.  Protection from local perturbations arises from non-local information storage, while the robustness of quantum gates is provided by path-independent braiding operations~\cite{Moore1991,Read2000, Ivanov2001,Kitaev2003,Chamon1997,DSarma2005,Stern2006,Bonderson2006,Nayak2008,Alicea2011}.  One way to build topological quantum computers involves end modes of topological superconducting nanowires known as Majorana bound states (MBSs)~\cite{Kitaev2001,Nayak2008,Beenakker2013,Alicea2012,Sau2010a,Alicea2010,Lutchyn2010,Oreg2010,Sau2010}. MBS are predicted to display non-Abelian braiding statistics in quasi-one dimensional networks. Signatures of MBSs have been reported in experiments such as zero bias conductance peaks in tunneling ~\cite{Mourik2012, Chen2017,Deng2016}. 

Josephson junctions formed between topological superconducting nanowires with a pair of MBS localized across the junction are predicted to exhibit a $4\pi$ Josephson effect~\cite{Kitaev2001,Marra2016}.
Evidence of the $4\pi$ Josephson effect has also been reported in microwave measurements on topological materials~\cite{Deacon2017, Wiedenmann2016}.

The future demonstration of non-abelian statistics will be implemented by braiding MBSs, which results in non-Abelian transformations through the Hilbert space of degenerate ground states~\cite{Ivanov2001,Alicea2011}.  The first braiding proposals involved interferometry of non-Abelian excitations hosted by fractional quantum Hall states~\cite{Chamon1997,DSarma2005,Stern2006,Bonderson2006,Nayak2008}.  
The next set of proposals suggested probing the non-Abelian statistics of MBSs hosted either in superconducting vortices or in chiral edge modes of 2D topological superconductors~\cite{Grosfeld2011, Mesaros2011, Sau2011a, Benjamin2010, Akhmerov2009, Fu2009, Sarma2006, Auslaender2008}.   
Following experimental reports of MBS in one-dimensional nanowires, a set of proposals has been developed based on nanowire tri-junctions by means of manipulating the chemical potential or charging energy~\cite{Alicea2011,Halperin2012, Hyart2013,Heck2012,Hassler2010,Hassler2011}. Proposals that do not make use of a tri-junction include measurement-based braiding using Majorana teleportation~\cite{Bonderson2009, Vijav2016}, and nanowire networks which once again use electrostatic gates to control the topological phase transition~\cite{Sau2011}.  Since none of the above schemes have been implemented experimentally, and every known scheme comes with limitations, it is important to further conceptualize braiding mechanisms in search for streamlined and complimentary approaches.

In this paper, we conceptualize a braiding scheme that uses the $4\pi$ topological Josephson effect to turn on and off couplings between neighboring MBSs across a tri-junction.  Furthermore, we couple the tri-junction to a Majorana-fluxonium qubit~\cite{Pekker2013} which is used to initialize and readout the quantum information.  Our device constitutes a fully functional topological qubit for testing the non-Abelian properties of topological superconductors.  One major advantage of our scheme is that flux control is already prevalent in superconducting information technology.  In fact, rapid single flux quantum devices, where classical information is encoded in a single quantum of flux, has been developed nearly to the point of commercialization \cite{Brock2001}.  Furthermore, the non-topological components of our device such as the fluxonium qubit have already been realized in experiment based on magnetic field-resilient elements compatible with Majorana physics\cite{Samkharadze2016,Hazard2018}.  It should be stressed that, although other flux controlled braiding devices have been proposed \cite{Hyart2013,Heck2012}, our device is fundamentally different in that we utilize flux to control the phase difference between Majorana nanowires, and couple-decouple Majoranas on adjacent wires via the 4$\pi$ Josephson effect,  instead of controlling the charging energy of Majorana islands. 

This paper is organized as follows: in section~\ref{TheBraidingCircuit} we discuss the circuitry used for braiding, in section~\ref{The4PiJosephsonBraidingProcedure} we  demonstrate how the braiding cuircut is used to flip the state of the topological qubit,  in section~\ref{TheFullQubitCircuit} we present the design for the full qubit including readout and initialization circuitry,  in section~\ref{InitializationAndReadout} we describe the procedure for initializtion and readout, in section~\ref{EffectOfFluxErrorsAndLowFrequencyNoiseOnBraiding} we discuss the effects of flux error and low frequency noise, and in section~\ref{ControllingThePhaseDifferenceViaExternalFlux} we show how the ability to control phase differences via the external magnetic flux depends on the inductance, capcitance, and Josephson energy of the device.  In section~\ref{Conclusion} we conclude.

\section{The Braiding Circuit}
\label{TheBraidingCircuit}
\begin{figure}[t]
\begin{center}
\includegraphics[width=\columnwidth]{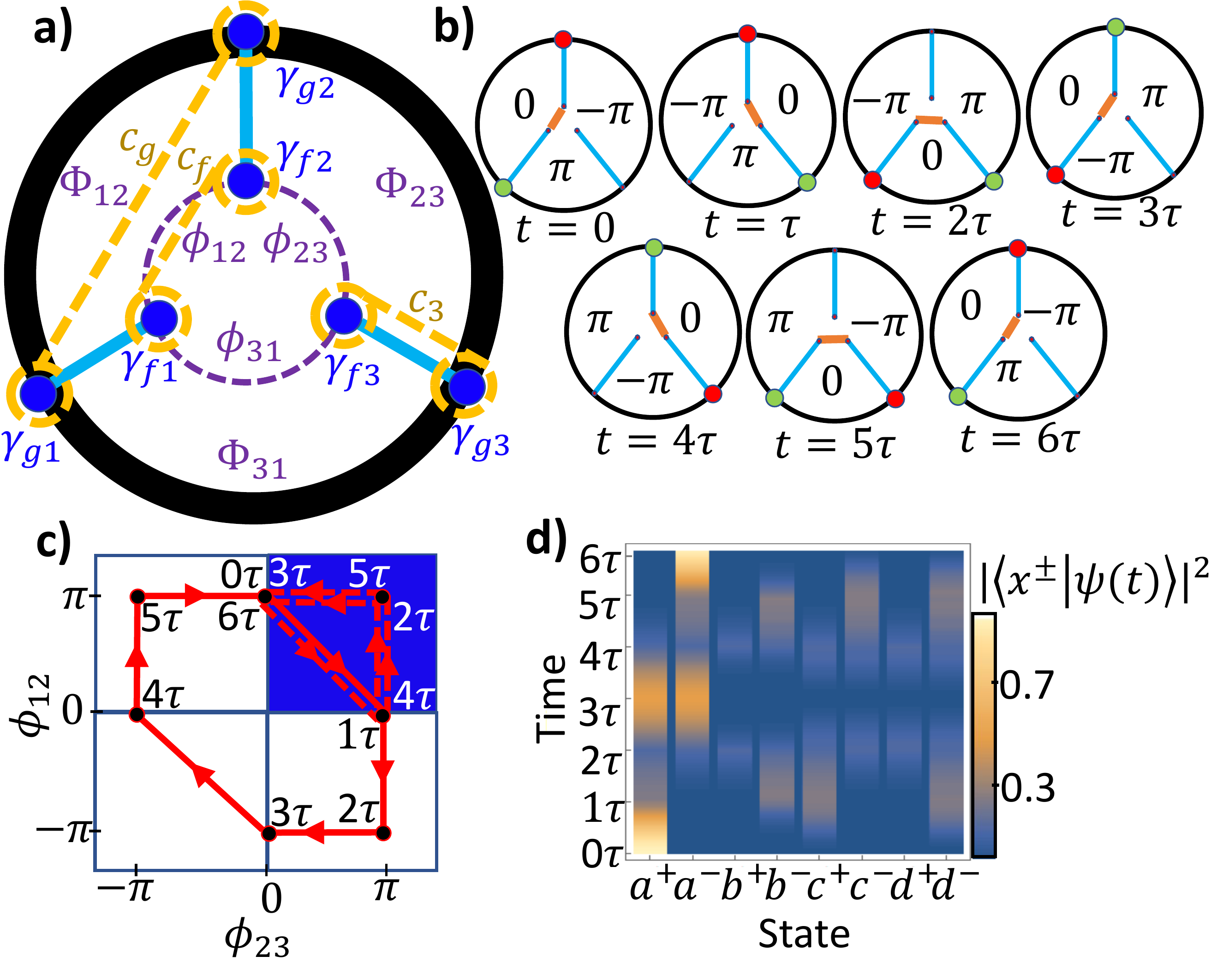}
\end{center}
\vspace{-2mm}

\caption{(a) Topological Josephson tri-junction ring.  The light blue segments are p-wave wires.  The MBSs are marked by blue dots and labeled $\gamma_n$.  The phase difference between superconductors $i$ and $j$ is labeled $\phi_{ij}$, while the external flux through the loop formed by superconductors $i$ and $j$ is labeled $\Phi_{ij}$.  The dotted yellow lines show the electron basis used in the text labeled $c_n$.  (b) Braiding sequence with the location of the MBSs (red and green dots) and the phase differences $\phi_{ij}$.  The inner MBS coupling is depicted by an orange bar. (c) The solid red line shows the braiding path that is required by the flux corral.  The Hilbert space trajectory along the solid path is equivalent to going twice around the dotted red triangle.  (d) The probability of being in a specific state $x^{\pm}\in\{a,b,c,d\}$ during the double braid process.}
\label{F1}
\vspace{-3mm}
\end{figure}

Figure~\ref{F1}(a) depicts the flux capacitor-shaped~\cite{Gale1985} $4\pi$-Josephson braiding circuit. The device consists of a conventional superconducting ring that is subdivided using one dimensional spinless p-wave topologically superconducting nanowires (referred to as `p-wave wires' in what follows).  The three nanowires come together at the center of the ring making a multiterminal topological Josephson tri-junction. The braiding ring hosts six MBSs, three at the tri-junction and three at the outer ends of the p-wave wires.  Using the $4\pi$ Josephson effect we can couple and uncouple the inner MBSs $\gamma_{f1}$, $\gamma_{f2}$, and $\gamma_{f3}$.  When the phases on any two arms of the junction are equal, the MBSs on those arms are coupled, when the two arms are out of phase by $\pi$ the MBSs are uncoupled.  These phase differences are tuned by threading external magnetic flux through the three loops formed by shorting the outer ends of the three nanowires through a non-topological superconductor ring.  To create MBSs it is necessary to apply in-plane magnetic fields of the order 0.1-1T oriented so that MBS exist in all arms of the junction. The ideal angle between crossed nanowires is 30 degrees \cite{Plissard2013} with the field oriented half-way between the nanowires.  Although the out of plane fields are too weak to drive topological transitions in the nanowires, care should be taken to avoid or compensate flux in the qubit loops due to the in-plane field. There has already been some work in this direction~\cite{Hart2016}.  In the future Zeeman splitting may be generated by local nanomagnets eliminating the need for large external in-plane magnetic fields.  The maximum coupling of the inner MBSs is given by the $4\pi$ Josephson energy scale $E_M$ which is determined by the tunneling amplitude between the inner MBSs.  For clarity of the presentation we focus on the case where all three $4\pi$ Josephson energies are the same and the trivial $2\pi$ Josephson effect is negligible. We note that the $2\pi$ Josephson effect can displace the phase differences at the tri-junction set by the external flux.  However, this can be easily avoided by ensuring that $E_J \ll E_L$.  This will be discussed in more detail in section \ref{ControllingThePhaseDifferenceViaExternalFlux}."

The length of the nanowire arms and their chemical potential is set so that two MBSs on the same arm (e.g. f1 and g1) have overlapping wave functions and hence their occupation acquires a small, but finite energy gap $\alpha$.  However, when two of the arms (e.g. 1 and 2) are coupled, the occupation energy gap for the outer two MBSs on those arms (g1 and g2) becomes $\alpha^2/E_M$. Therefore, on timescales of $1/\alpha<t/2\pi\hbar<E_m/\alpha^2$, the level repulsion of the outer two MBSs is negligible and their energies are not distinguishable from zero.  In this case, we will refer to the outer MBSs as Majorana zero modes (MZMs) and will use these for braiding since they are topologically protected during the aforementioned timescale.  We initialize the three fluxes $\Phi_{12},\Phi_{23},\Phi_{31}$ so that $\gamma_{f1}$ and $\gamma_{f2}$ are coupled which means $\gamma_{g1}$ and $\gamma_{g2}$ are the MZMs at the start of braiding while $\gamma_{f3}$ and $\gamma_{g3}$ are auxiliary MBSs. For this reason, it is useful to work in the complex fermion basis $c_g=1/2(\gamma_{g1}+i\gamma_{g2})$, $c_f=1/2(\gamma_{f1}+i\gamma_{f2})$, and $c_3=1/2(\gamma_{f3}+i\gamma_{g3})$ as labeled in Fig.~\ref{F1}(a).  

%%%%%%%%%%%%%%%%%%%%%%%%%%%%%%%%
%%%%%%%%%%%%%%%%%%%%%%%%%%%%
%%%%%%%%%%%%%%%%%%%%%%	
%%%%%%%%%%%%%%%%%%%%%%%%%%%%%%%%

We will focus on the case of rigid superconductivity, in which the phases $\phi_{12}$, $\phi_{23}$, and $\phi_{31}$ at the tri-junction do not undergo quantum fluctuations. The rigidity condition is satisfied when the inductive energy $E_L$ of the outer superconducting ring dominates the charging energy $E_C$, and the Majorana Josephson energy $E_M$ at the tri-junction (If either $E_L/E_M$ or $E_L/E_C$ is not large enough, the phase differences can delocalize).  Furthermore, we shall restrict ourselves to the ``flux corral" condition in which the total flux through all three loops is zero $\Phi_{12}+\Phi_{23}+\Phi_{31}=0$. With these restrictions, the external fluxes directly control the phases at the tri-junction: $\phi_{ij}=\Phi_{ij}$ and the Majorana bound state Hamiltonian for the braidonium ring is:
\begin{equation}
\label{HMm}
\begin{split}
&H^t_{M}=iE_M\sum_{i=1}^3{\gamma_{fi}\gamma_{fi+1}}\cos\left(\frac{\Phi_{i,i+1}}{2}\right)
+i\alpha\sum_{i=1}^3\gamma_{fi}\gamma_{gi}
\end{split}
\end{equation}
where the index $i$ is defined modulo 3 (i.e. $i=3+1=1$).

\section{The 4$\pi$ Josephson braiding procedure}
\label{The4PiJosephsonBraidingProcedure}

The braiding process consists of quasi-adiabatical transitions between flux configurations in which the tri-junction has one coupling turned on (say $\Phi_{12}=0$) and the other two off ($\Phi_{23}=\pi$ and $\Phi_{31}=-\pi$).  The steps of the braiding procedure are depicted in Fig.~\ref{F1}(b)-(c).  Both panels show the external flux settings at the end of each step of the braiding process.  Panel (b) shows the position of the two Majoranas being braided (red and green dots) during the process, as well as the coupling between the tri-junction arms (orange bar). During each time step $\tau$, the coupling between one pair of MBSs is slowly (adiabatically) turned off while the coupling between another pair is slowly turned on.  In the first step, for example, the external flux $\Phi_{12}$ is tunned from $0$ to $-\pi$ which turns off the coupling between $\gamma_{f1}$ and $\gamma_{f2}$, at the same time $\Phi_{23}$ is tuned from $-\pi$ to $0$ which turns on the coupling between $\gamma_{f2}$ and $\gamma_{f3}$.  As these couplings change, the green Majorana loses weight in $\gamma_{g1}$ and gains weight in $\gamma_{g3}$.  At the end of the step it has moved completely from the left arm to the right arm.  The entire process braids the two Majoranas twice around each other.  Panel (c) shows the braiding path through flux space.  Although the fluxes at $t=3\tau$ and $t=0$ are different, a single exchange has still occurred at $t=3\tau$ as the Hamiltonian (Eq.~\ref{HMm}) is an even function of the external flux parameters.  

After the double exchange, $c^{\dagger}_g\rightarrow-c^{\dagger}_g$.  We will use this change of sign to readout the state of our topological qubit.  In order to detect the change of sign, however, we need to interfere linear combinations of $\ket{n_g=0}$ and $\ket{n_g=1}$ which are in different parity sectors.  Since parity is conserved, the only way to form linear combinations of these states is to introduce an additional pair of MZMs, $\gamma_{r1}$ and $\gamma_{r2}$ which act as a parity reservoir elsewhere in the device (see discussion below). By coupling one of these reservoir MZMs ($\gamma_{r1}$ or $\gamma_{r2}$) to one of the initial  MZMs ($\gamma_{g1}$ or $\gamma_{g2}$) we can interfere the two occupation states ($\ket{n_g=0,1}$) without breaking parity conservation.  We describe the state of these MZMs using the complex fermion basis $\ket{n_g,n_f,n_3,n_r}$ where $n_i$ is the occupation of the $i^{th}$ electron ($i\in\{g,f,3,r\}$).  Considering only even parity states, we have:
\begin{eqnarray}
\label{states}
&\ket{a^+}=\ket{0000}+\ket{1001} \quad \quad &\ket{a^-}=\ket{0000}-\ket{1001}
\nonumber
\\
&\ket{b^+}=\ket{0101}+\ket{1100} \quad \quad &\ket{b^-}=\ket{0101}-\ket{1100}
\\
&\ket{c^+}=\ket{0110}+\ket{1111} \quad \quad &\ket{c^-}=\ket{0110}-\ket{1111}
\nonumber
\\
&\ket{d^+}=\ket{0011}+\ket{1010} \quad \quad &\ket{d^-}=\ket{0011}-\ket{1010}
\nonumber
\end{eqnarray}
Each of these states is a linear combination of $\ket{n_g=0}$ and $\ket{n_g=1}$ so that the braiding procedure takes all (+) type states (i.e. $\ket{a^+}$) to (-) type states (i.e. $\ket{a^-}$).  A topological qubit can be formed out of any linear combination of (+) type states and the corresponding linear combination of (-) type states.  In Fig.~\ref{F1}(d) we show the probability that the state of the tri-junction $\psi(t)$ is in a particular occupation state during the braiding procedure.  Here, we show the case where the qubit is formed by $\ket{a^+}$ and $\ket{a^-}$ so that $\psi(0)=\ket{a^+}$ and $\psi(6\tau)=\ket{a^-}$.

%%%%%%%%%%%%%%%%%%%%%%%%%%%%%%%%
%%%%%%%%%%%%%%%%%%%%%%%%%%%%
\begin{figure}[t]
\begin{center}
\includegraphics[width=0.5\textwidth]{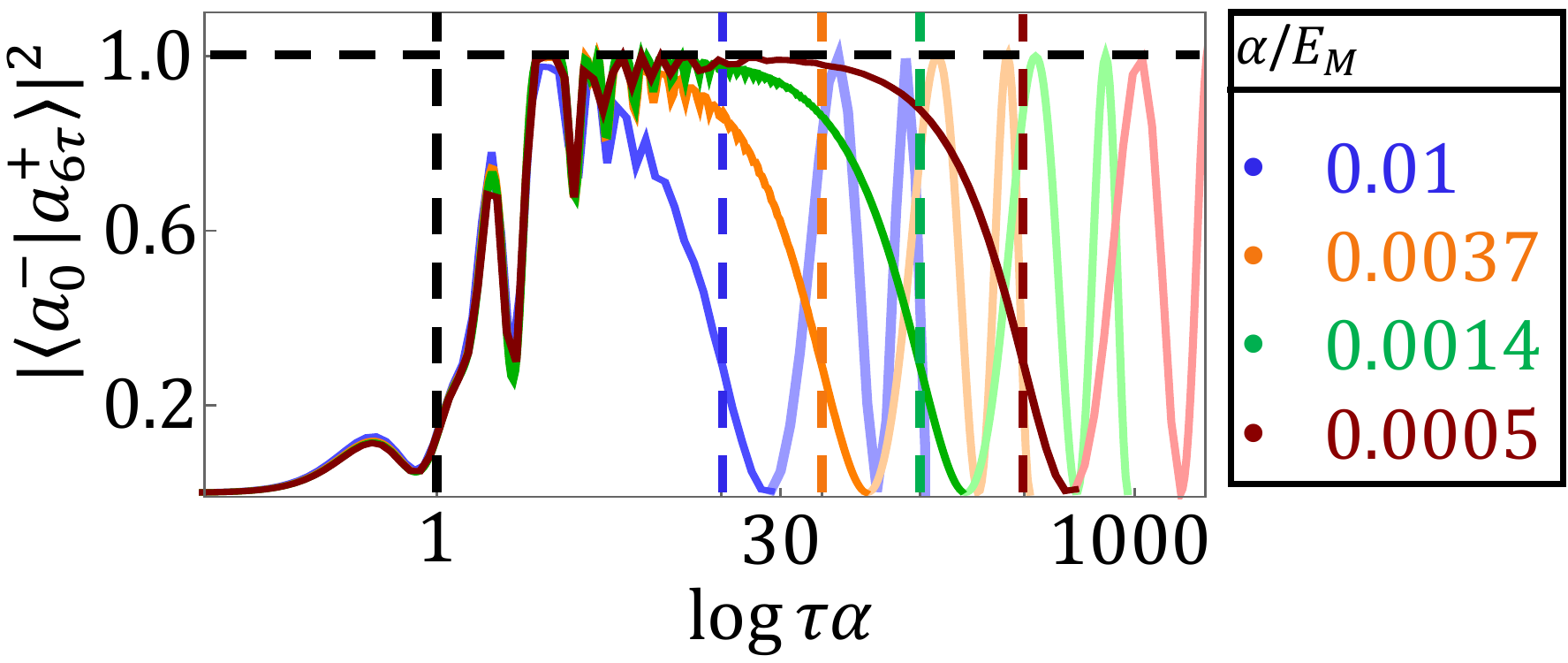}
\end{center}
\vspace{-2mm}

\caption{Probability of being in the expected state $\ket{a^-}$ after the double braid plotted against the log of the duration $\tau$ of each step (in units of $1/\alpha$) for different value of $\alpha$.   The horizontal dotted black line marks perfect braiding.  The vertical dashed lines mark the time scales $\tau=1/\alpha$ (black) and $\tau=E_M/6\alpha^2$ (color corresponding to curve).}
\label{F3}
\vspace{-3mm}
\end{figure}
%%%%%%%%%%%%%%%%%%%%%%	
%%%%%%%%%%%%%%%%%%%%%%%%%%%%%%%%

The time scale of the braiding procedure has to be slow enough to be adiabatic but fast enough that we can treat MBS as true zero modes (MZM).  It takes approximately $t\approx 2\pi\hbar/\alpha$ for a Majorana to travel from one end of the nanowire to the other.  Therefore, each step of the braiding procedure $\tau$ must last longer than $2\pi\hbar/\alpha$.  However, the energy gap between $\ket{n_g=0}$ and $\ket{n_g=1}$ is on the order of $\alpha^2/E_M$, therefore 
if braiding lasts longer than $2\pi\hbar E_M/\alpha^2$ we will start to lose coherence.  Figure~\ref{F3} shows the fidelity of the double braiding procedure.
As seen in Fig.~\ref{F3}, the fidelity plateaus near unity when $1 \lesssim \tau \alpha \lesssim E_M/6\alpha$ (the factor of six comes from the fact that the entire procedure takes $6\tau$ to complete).  The further these limiting timescales are from each other (i.e. the larger $E_M/\alpha$), the greater the range of acceptable time scales for braiding.  However, the smaller $\alpha$, the slower the process has to be, thus, the coherence of the device must be longer.  One will want to make $\alpha$ as small as allowed by the fastest decoherence process, e.g. the quasi-particle poisoning timescale~\cite{Pop2014,Higginbotham2015}. Based on the induced gap of the p-wave wires, we expect $E_M\sim 0.1$~meV~\cite{Chen2017} which puts $\tau\sim 100$~ns.

%%%%%%%%%%%%%%%%%%%%%%%%%%%%%%%%
%%%%%%%%%%%%%%%%%%%%%%%%%%%%
\begin{figure}[t]
\begin{center}
\includegraphics[width=0.8\columnwidth]{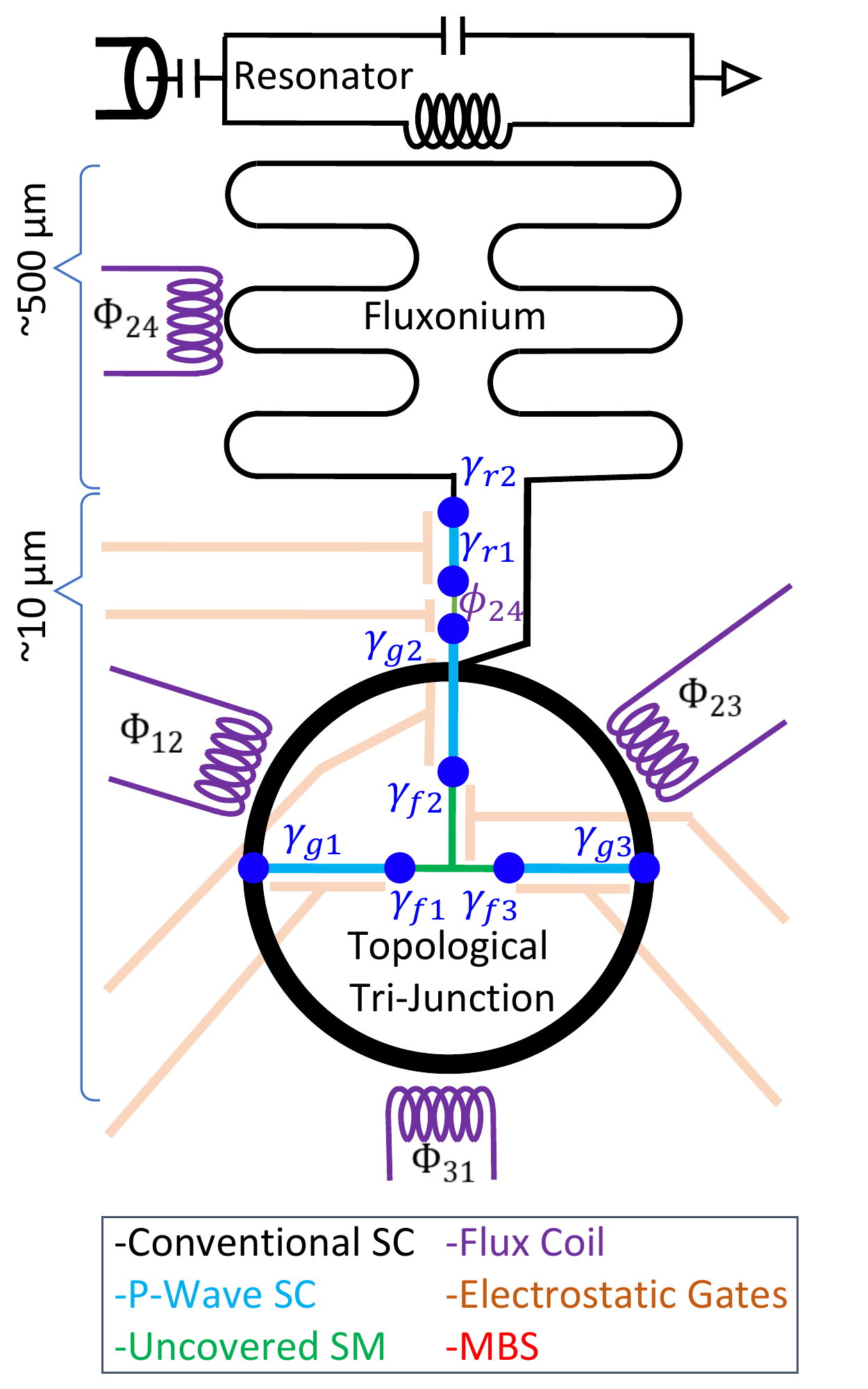}
\end{center}
\vspace{-2mm}

\caption{
Full braidonium device. A fluxonium loop is coupled to the transmission line via a microwave resonator. The fluxonium loop is coupled to the tri-junction loop through the p-wave wire (light blue lines).  External flux (fast) through the fluxonium loop and each of the three sections of the braiding ring is controlled by on-chip coils.  Several electrostatic gates (slow) are used to tune the superconducting nanowires into the topological regime.  The location of MBSs are marked by blue dots.  Uncovered regions of semiconductor nanowire used for the Josephson junctions are depicted as green lines.}
\label{F4}
\vspace{-3mm}
\end{figure}
%%%%%%%%%%%%%%%%%%%%%%	
%%%%%%%%%%%%%%%%%%%%%%%%%%%%%%%%

\section{The full qubit circuit}
\label{TheFullQubitCircuit}
In the full braidonium qubit, in order to initialize and readout the topological qubit, we propose to integrate it with an ancilla Majorana-fluxonium qubit previously proposed in Ref.~[\onlinecite{Pekker2013}]. The fluxonium will be coupled to a transmission line via a microwave resonator made of magnetic field-resilient superconductor, e.g. NbTiN. A schematic of the entire system is depicted in Fig.~\ref{F4}. 
The coupling between the fluxonium and the topological qubit will also proceed via the $4\pi$ Josephson effect which couples $\gamma_{r1}$ and $\gamma_{g2}$ to the phase $\phi_{24}$ across the Josephson junction of the fluxonium. The coupling between $\gamma_{r1}$ and $\gamma_{r2}$ is much smaller than $\alpha$ so that they can be treated as MZMs during braiding.  The external flux is controlled by four inductance coils (one for the fluxonium qubit and three for the tri-junction).  A number of electrostatic gates could be used to tune the chemical potential in various sections of the nanowires.  However, once the potentials on the gates are set, they are not changed in the course of braiding.  Unlike the tri-junction ring, we want the fluxonium to undergo phase slips and so the phase difference $\phi_{24}$ is not necessarily equal to $\Phi_{24}$ the external flux through the fluxonium.  Therefore we have to include the non-topological part of the fluxonium Hamiltonian 
\begin{equation}
\label{HFloop}
H^{l}_{NT}(\phi_{24},\Phi_{24})=E^l_C\partial^2_{\phi_{24}}-E^l_L(\phi_{24}-\Phi_{24})^2-E^l_J\cos(\phi_{24})
\end{equation}
where the fluxonium loop parameters are $E^l_J$ the Josephson energy, $E_C^l=e^2/2C^l$ the charging energy, and $E_L^l=\hbar^2/4e^2L^l$ the inductive energy, with $C^l$ being the capacitance and $L^l$ the inductance of the fluxonium loop.  Phase slips occur when $E^l_C \gg E^l_L$.  The Majorana part of the the full Hamiltonian is 
\begin{equation}
\label{HM}
H_M=H_M^t+i\alpha_r\gamma_{r1}\gamma_{r2}+i E_m^l\cos\left(\frac{\phi_{24}}{2}\right)\gamma_{g2}\gamma_{r1}
\end{equation}
where $E^l_M$ is the Majorana Josephson energy for the fluxonium loop.  

When $E^l_J\sim E^l_m\gg E^l_C,E^l_L$, the low energy spectrum  of the non-topological Hamiltonian for the loop is a harmonic oscillator trapped in the potential wells formed by the cosine functions  \cite{Pekker2013}. We will use the first two flux quanta occupation states ($\ket{n_{\phi_{24}}=0}$ and $\ket{n_{\phi_{24}}=1}$) for initialization/readout.  The full Hamiltonian $H=H^l_{NT}+H_M$ can be solved numerically as we will now show. 

We will rewrite Eq. \ref{HM} in the complex fermion basis.  Let us start with Eq.~\ref{HMm}.  We take $c^{\dagger}_g=(\gamma_{g1}-i\gamma_{g2})/2$ to be the creation operator for the electron associated with the Majoranas that will be braided.  Then there are two auxiliary electrons $c^{\dagger}_{f}=(\gamma_{f1}-i\gamma_{f2})/2$ and $c^{\dagger}_3=(\gamma_{f3}-i\gamma_{g3})/2$ which are not involved in the process but which are nevertheless present.  Using this notation we can write the Majorana operators in terms of electron operators.
\begin{equation}
\begin{split}
&\gamma_{f1}=c^{\dagger}_f+c_f\quad\quad\gamma_{f2}=i(c^{\dagger}_f-c_f)
\\
&\gamma_{g1}=c^{\dagger}_g+c_g\quad\quad\gamma_{g2}=i(c^{\dagger}_g-c_g)
\\
&\gamma_{f3}=c^{\dagger}_3+c_3\quad\quad\gamma_{g3}=i(c^{\dagger}_3-c_3)
\end{split}
\end{equation}
In terms of these electron operators, the tri-junction Hamiltonian, Eq.~\ref{HMm}, becomes
\begin{equation}
\label{HMe}
\begin{split}
&H_M^t(\Phi_{12},\Phi_{23})=\alpha \left( n_3-\frac{1}{2} \right)+2i\alpha \left(c^{\dagger}_f c_g+c_f c^{\dagger}_g \right)
\\
&+E_M \left( n_f-\frac{1}{2} \right) \cos \left( \frac{\Phi_{12}}{2} \right)
\\
&-E_M \left( c^{\dagger}_fc^{\dagger}_3+c^{\dagger}_f c_3 \right) \left( \cos\left(\frac{\Phi_{23}}{2}\right)+i \cos\left(\frac{\Phi_{31}}{2}\right) \right)
\\
&+E_M \left(c_f c^{\dagger}_3+c_f c_3 \right) \left( \cos\left(\frac{\Phi_{23}}{2}\right)-i \cos\left(\frac{\Phi_{31}}{2}\right) \right),
\end{split}
\end{equation}
where $n_f=c^{\dagger}_f c_f$ and $n_3=c^{\dagger}_3 c_3$ are number operators.  Defining the third number operator, $n_g= c^{\dagger}_g c_g$, we can now write the Hamiltonian in the basis $\ket{n_f,n_g,n_3}$
\begin{equation}
 \label{HMs}
 \begin{split}
 &H_{M,e}^{t}(\Phi_{12},\Phi_{23})=
 \\
 &E_M\left( \cos\left(\frac{\Phi_{12}}{2}\right)\sigma^z_f+\cos\left(\frac{\Phi_{23}}{2}\right)\sigma^z_f\sigma^x_3+\cos\left(\frac{\Phi_{31}}{2}\right)\sigma^y_3 \right)
 \\
 &+\alpha\left(\sigma^z_3+2\sigma^y_f\delta_{n_3=1}\right),
 \\
 &H_{M,o}^{t}(\Phi_{12},\Phi_{23})=
 \\
 &E_M\left( \cos\left(\frac{\Phi_{12}}{2}\right)\sigma^z_f+\cos\left(\frac{\Phi_{23}}{2}\right)\sigma^z_f\sigma^x_3+\cos\left(\frac{\Phi_{31}}{2}\right)\sigma^y_3 \right)
 \\
 &+\alpha \left(\sigma^z_3+2\sigma^y_f\delta_{n_3=0} \right),
 \end{split}
 \end{equation}
where $H^t_{M,e}$ and $H^t_{M,o}$ correspond to the even and odd sectors of $\braket{n_f,n_g,n_3|H^t_M| n^{\prime}_f,n^{\prime}_g,n^{\prime}_3}$ and $\sigma^i_a$ is the $i^{th}$ Pauli matrix acting on the $a^{th}$ electron occupation basis $\ket{n_a}$.  Notice that, by separating the Hamiltonian into even and odd parity, we no longer need all three electron occupations to describe the state.  Two of the occupation numbers and the specification of either even or odd parity is enough.  We have made the choice to remove reference to the $\ket{n_g}$ occupation.  Now we can write the full Majorana Hamiltonian, Eq.~\ref{HM}, in terms of Pauli matrices just as we did for Eq~\ref{HMm}.  We use the basis $\ket{n_f,n_3}\bigotimes\ket{n_r}$; where $c^{\dagger}_r=1/2(\gamma_{r1}-i\gamma_{r2})$. We obtain,
\begin{equation}
\begin{split}
&H_M(\Phi_{12},\Phi_{23},\phi_{24})=
\\
&E_M\left( \cos\left(\frac{\Phi_{12}}{2}\right)\sigma^z_f+\cos\left(\frac{\Phi_{23}}{2}\right)\sigma^z_f\sigma^x_3+\cos\left(\frac{\Phi_{31}}{2}\right)\sigma^y_3 \right)
\\
&+\alpha\left(\sigma^z_3+2\sigma^y_f\delta_{n_3+n_r=1}\right)+E_M^l\cos\left(\frac{\phi_{24}}{2}\right)\sigma^y_r+\alpha_r\sigma_r^z,
\end{split}
\end{equation}
where $\sigma_r^i$ is the i-component Pauli matrix acting on the reservoir occupation basis $\ket{n_r}$. The parity reservoir couples the even and odd parity sectors of the tri-junction, however, total parity is still separable.  Here we have written only the even parity sector which is used throughout the text.  Odd parity is very similar. 

\begin{figure}[t]
\begin{center}
\includegraphics[width=\columnwidth]{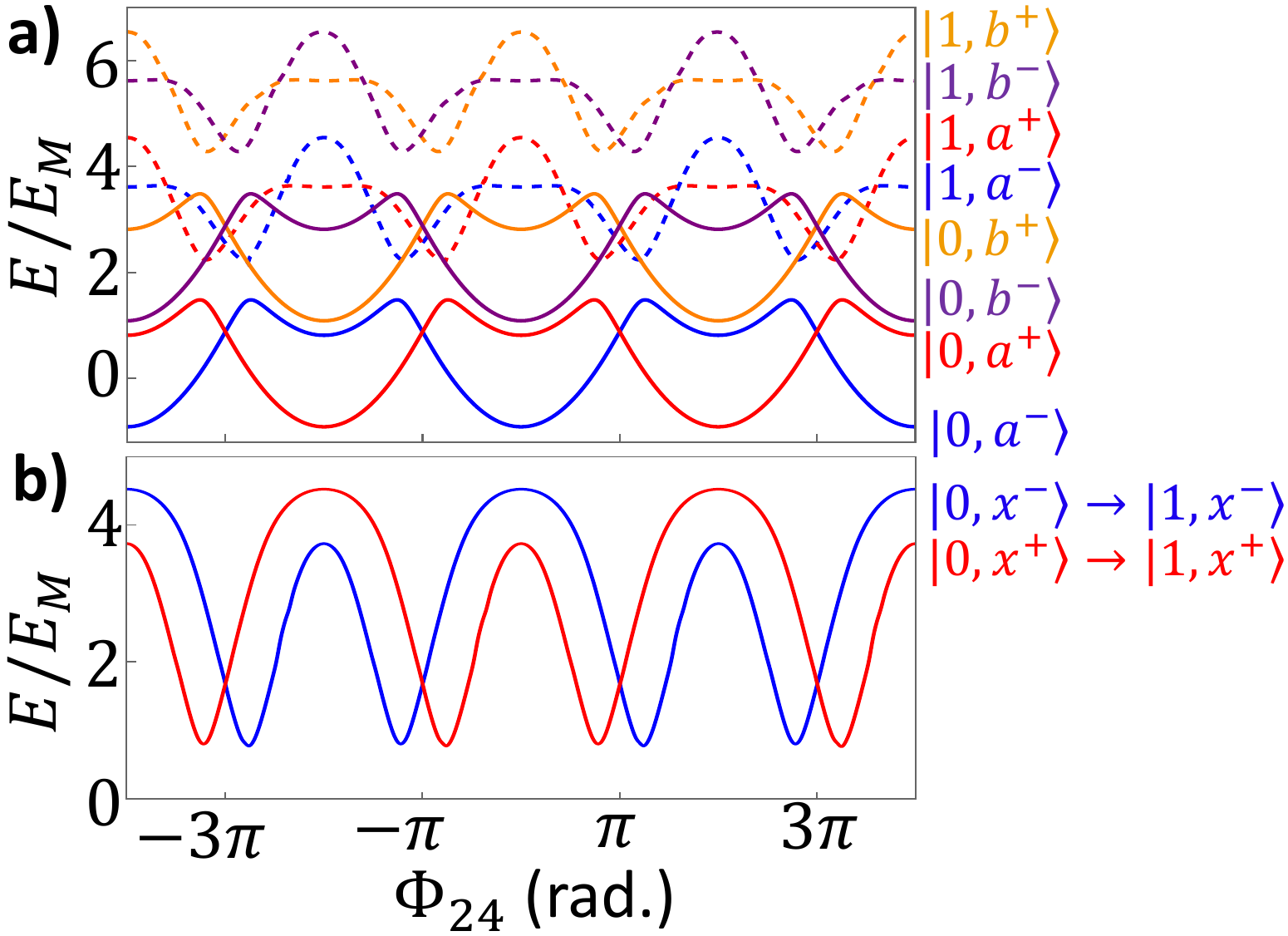}
\end{center}
\vspace{-2mm}

\caption{(a) Energy levels as a function of  $\Phi_{24}$.  The color of the curve reflects its electron occupation.  Each curve represents two nearly degenerate fermion occupations.  Red is both  $\ket{a^+}$ and $\ket{d^+}$, blue is $\ket{a^-}$ and $\ket{d^-}$, orange is $\ket{b^+}$ and $\ket{c^+}$, and purple is $\ket{b^-}$ and $\ket{c^-}$.  The dashed lines have the same color scheme as the solid lines, but the fluxonium is in the first excited state.  (b) Energy difference between the states $\ket{0,x^-}$ and $\ket{1,x^-}$ (blue) and states $\ket{0,x^+}$ and $\ket{1,x^+}$ (red). 
} 
\label{F5}
\vspace{-3mm}
\end{figure}

In order to finish preparing the full Hamiltonian for numerical calculations, we need to deal with the quantum phase $\phi_{24}$.  With a change of variables $\phi_{24}\rightarrow s\phi_{24}+\Phi_{24}$, where $s=(E_C/E_L)^{1/4}$, we can write the non-topological part of the fluxonium Hamiltonian (Eq.~\ref{HFloop}) in the harmonic oscillator basis,
\begin{equation}
\braket{\phi_{24}|n_{\phi_{24}}}=\frac{1}{\sqrt{2^{n}n!\sqrt{\pi}}}e^{-\frac{\phi^2}{2}}H_{n_{\phi_{24}}}(\phi_{24})
\end{equation}
where $H_{n}(\phi)$ are the Hermite polynomials.  In this basis we have,
\begin{equation}
\label{HFloopphi}
\begin{split}
&\braket{n_{\phi_{24}}|H^l_{NT}(\Phi_{24})|n_{\phi_{24}}+m_{\phi_{24}}}=\omega_l \left( n_{\phi_{24}}+\frac{1}{2} \right)\delta_{m_{\phi_{24}},0}
\\
&+E_J^l \left( C_{n_{\phi_{24}},m_{\phi_{24}}}(s)\cos(\Phi_{24})+S_{n_{\phi_{24}},m_{\phi_{24}}}(s)\sin(\Phi_{24})  \right)
\end{split}
\end{equation}
where $\omega_l=\sqrt{E_C^l E_L^l}$ and 
\begin{equation}
\begin{split}
&C_{n,m}(s)=
\\
&(-1)^{m/2}\frac{\sqrt{2^n n!}s^{m+1}e^{-b^2/4}}{\sqrt{2^{n+m}(n+m)!}}L^m_n\left(\frac{s^2}{2}\right)\delta_{(\text{Mod}_2(m),0)},
\\
&S_{n,m}(s)=
\\
&(-1)^{(m-1)/2}\frac{\sqrt{2^n n!}s^{m+1}e^{-b^2/4}}{\sqrt{2^{n+m}(n+m)!}}L^m_n\left(\frac{s^2}{2}\right)\delta_{(\text{Mod}_2(m),1)},
\end{split}
\end{equation}
in which $L^n_m(b)$ are the Laguerre polynomials and $\delta_{\text{Mod}_2(m),0}$ is zero unless $m$ is even while $\delta_{\text{Mod}_2(m),1}$ is zero unless $m$ is odd.  This Hamiltonian can be solved numerically, after suitably truncating the Hilbert space. Similarly, we can write the Majorana Hamiltonian in the harmonic oscillator basis,
\begin{equation}
\label{HMphi}
\begin{split}
&\bra{n_{\phi_{24}}}  H_{M}(\Phi_{12},\Phi_{23},\Phi_{31},\Phi_{24})  \ket{n_{\phi_{24}}+m_{\phi_{24}}}= 
\\
&\alpha (\sigma^z_3 +2\sigma^y_f\delta_{n_3+n_g,1})\delta_{m_{\phi_{24}},0}+E_M \cos\left(\frac{\Phi_{12}}{2}\right)\sigma^z_f\delta_{m_{\phi_{24}},0}
\\
&+E_M \cos\left(\frac{\Phi_{23}}{2}\right)\sigma^z_f\sigma^x_3\delta_{m_{\phi_{24}},0}+E_M\cos\left(\frac{\Phi_{31}}{2}\right)\sigma^y_3 \delta_{m_{\phi_{24}},0}
\\
&+E^l_{M}C_{n_{\phi_{24}},m_{\phi_{24}}}\left(\frac{s}{2}\right)\cos\left(\frac{\Phi_{24}}{2}\right)\sigma^y_l
\\
&-E^l_{M}S_{n_{\phi_{24}},m_{\phi_{24}}}\left(\frac{s}{2}\right)\sin\left(\frac{\Phi_{24}}{2}\right)\sigma^y_l
\end{split}
\end{equation}
Together Eq.~\ref{HFloopphi} and Eq.~\ref{HMphi} define the full Hamiltonian whose energy levels are plotted in Fig. \ref{F5}.  In constructing the figure, we set $E_C^l=5~E_M$, $E_L^l=0.25~E_M$, $E_J^l=2.5~E_M$, and $E_M^l=E_M$.  With these parameters the energy levels that we plot are fully converged when we truncate the Hilbert space to the lowest ten Harmonic oscillator levels in the flux basis $(n_{\phi_{24}}=1,...,10)$.

%%%%%%%%%%%%%%%%%%%%%%%%%%%%%%%%
%%%%%%%%%%%%%%%%%%%%%%%%%%%%
%%%%%%%%%%%%%%%%%%%%%%	
%%%%%%%%%%%%%%%%%%%%%%%%%%%%%%%%

\section{Initialization and readout}
\label{InitializationAndReadout}

In order to initialize and readout the fermion state, we use the shift of the fluxonium frequency in response to the change in occupation of fermion $c_r$ (see Fig~\ref{F5}).  Specifically, this shift allows us to distinguish between (+) and (-) type states using conventional quantum electrodynamic techniques \cite{Blais2004,Geerlings2013}. This is accomplished by (1) driving the system at a frequency where the cavity response distinguishes the fluxonium state and projects the fluxonium into a specific state; and (2) using the fact that the energy separation of the fluxonium levels depends on the fermion state to selectively drive fluxonium transitions. 

We propose to drive the system using a resonator as depicted in Fig~\ref{F4}.  We model the resonator and its coupling to the fluxonium qubit using the dispersive Hamiltonian:
\begin{equation}
H_R=\hbar\omega_0a^{\dagger}a+\frac{1}{2}\hbar\chi a^{\dagger}a(\sigma_{\phi_{24}}^z-1)+H
\end{equation}
where $H=H^{l}_{NT}+H_M$ is the full Hamiltonian, $a^{\dagger}$, $a$ are the creation and annihilation operators for the resonator and $\sigma_{\phi_{24}}^z$ is the z-component Pauli matrix in the basis of the first two fluxonium levels which form the fluxonium qubit, $\omega_0$ is the bare resonator frequency, and $\chi$ is the dispersive shift of the resonator by the fluxonium.  When the fluxonium is in its ground state, the resonator has a resonance peak at $\omega_0$. If, however, the fluxonium is in its first excited state then the resonance peak is shifted to $\omega_0+\chi$.  
The state of the fluxonium ($\ket{n_{\phi_{24}}=0}$ or $\ket{n_{\phi_{24}}=1}$) can be measured by  driving the resonator at either $\omega_0$ or $\omega_0+\chi$ and observing whether the cavity transmits or reflects the drive photons.  In order to collapse the system into a particular fluxonium state, we perform  multiple measurements until the desired outcome is achieved.  
At this point the fermion state remains unknown.  In order to observe whether the fermion state is (+) or (-) type, we selectively flip the state of the fluxonium.  Since fluxonium shows different resonance frequencies (i.e. different $\chi$) for (+) and (-) type states, a selective $\pi$ pulse will flip the state of the fluxonium only if it is specific to the particular fermion state \cite{Geerlings2013}.  Then we measure the state of the fluxonium again.  If it has changed state then we know that it is in the desired fermion state (i.e. $|a^+\rangle$, $|b^+\rangle$, $|c^+\rangle$, or $|d^+\rangle$). Otherwise, we reset and repeat the entire process until we get the desired fermion state. 

We pause to point out that our initialization procedure does not have control over which of the (+) type states we obtain. However, as we show below, the braiding operation works as long as the initial state is in the (+) sector.  

The explicit steps of the initialization procedure to project onto the ground state of the fluxonium and a $(+)$ type fermion state are outlined below:

1) Measure the fluxonium by driving the cavity at the excited state frequency.  Reset and repeat until the fluxonium is in its excited state.

2) Send in a selective $\pi$ pulse which will flip the state of the fluxonium only if the system is in a $(+)$ type fermion state.

3) Measure the state of the fluxonium.  If it is in the ground state then the procedure is complete.  Otherwise restart at step one.

%%%%%%%%%%%%%%%%%%%%%%%%%%%%%%%%
%%%%%%%%%%%%%%%%%%%%%%%%%%%%
\begin{figure}[t]
\begin{center}
\includegraphics[width=\columnwidth]{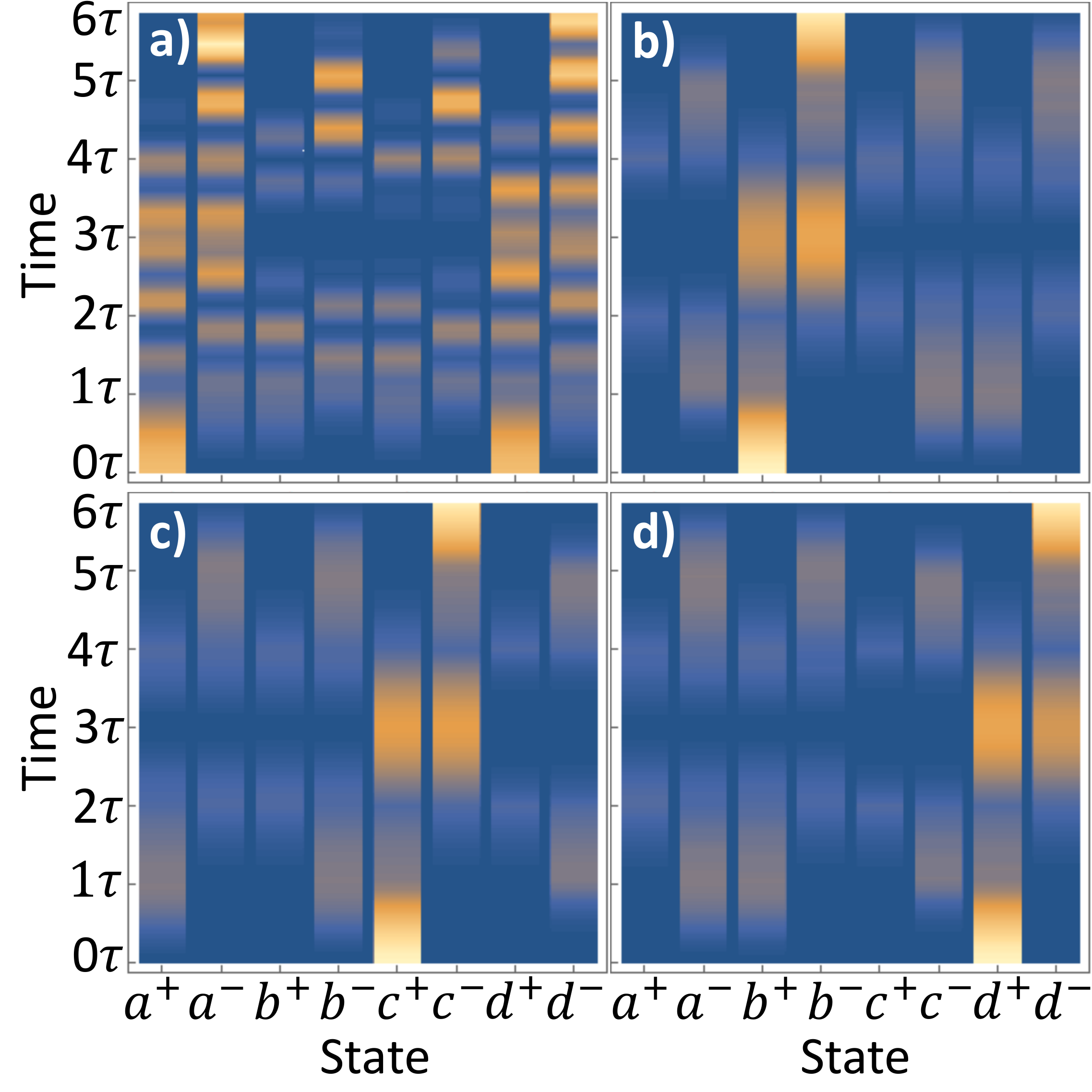}
\end{center}
\vspace{-3mm}
\caption{Probability of being in specific state during the double braid process for different initial states. For panel (a), we starts in the state $1/\sqrt{2}(\ket{a^+}+\ket{d^+})$ while in panels (b-d) we start in eigenstates $\ket{b^+}$, $\ket{c^+}$, and $\ket{d^+}$ respectively.  A similar plot where the initial state is $\ket{a^+}$ is shown in Fig~\ref{F1}(d).}
\label{F7}
\vspace{3mm}
\end{figure}
%%%%%%%%%%%%%%%%%%%%%%	
%%%%%%%%%%%%%%%%%%%%%%%%%%%%%%%%

Figure~\ref{F5}(a) shows energy levels for the full braidonium Hamiltonian as a function of $\Phi_{24}$. Indeed, we observe that the energy difference between the ground and excited levels depends on the whether the fermion state is a (+) or (-) type state (see~Figure~\ref{F5}(b)) as long as $\Phi_{24} \neq \pm\pi,\pm3\pi, \dots$.  In order to observe braiding, we set $\Phi_{24}$ to a generic value, and initialize the braidonium into the state described by the density matrix
\begin{align}
\label{rho}
\rho=\ket{n_{\phi_{24}}=0}\bra{n_{\phi_{24}}=0}\otimes\sum_x\sum_ys_{x,y}\ket{x^+}\bra{y^+}
\end{align}
where $x,y\in\{a,b,c,d\}$ label the fermion state and $s_{x,y}$ are arbitrary constants. Next, we tune to the decoupling point $\Phi_{24}=\pi$, and perform the double braiding procedure.  This procedure takes any $\ket{x^+}\rightarrow\ket{x^-}$ as seen in Fig.~\ref{F7}. Therefore, the density matrix becomes 
\begin{align}
&\rho\rightarrow \rho=\ket{n_{\phi_{24}}=0}\bra{n_{\phi_{24}}=0}\otimes\sum_x\sum_y s_{x,y}\ket{x^-}\bra{y^-}
\end{align}
To determine whether the braiding process has been successful (i.e. readout), we tune $\Phi_{24}$ to a generic point and apply a $\pi$ pulse corresponding to the $\ket{0,x^-} \rightarrow \ket{1,x^-}$ transition frequency (blue line in Fig.~\ref{F5}(b)). Successful braiding is indicated by fluxonium being promoted to the excited state.

\section{Effect of Flux Errors and Low Frequency Noise on Braiding}
\label{EffectOfFluxErrorsAndLowFrequencyNoiseOnBraiding}

\begin{figure}[t]
\begin{center}
\includegraphics[width=\columnwidth]{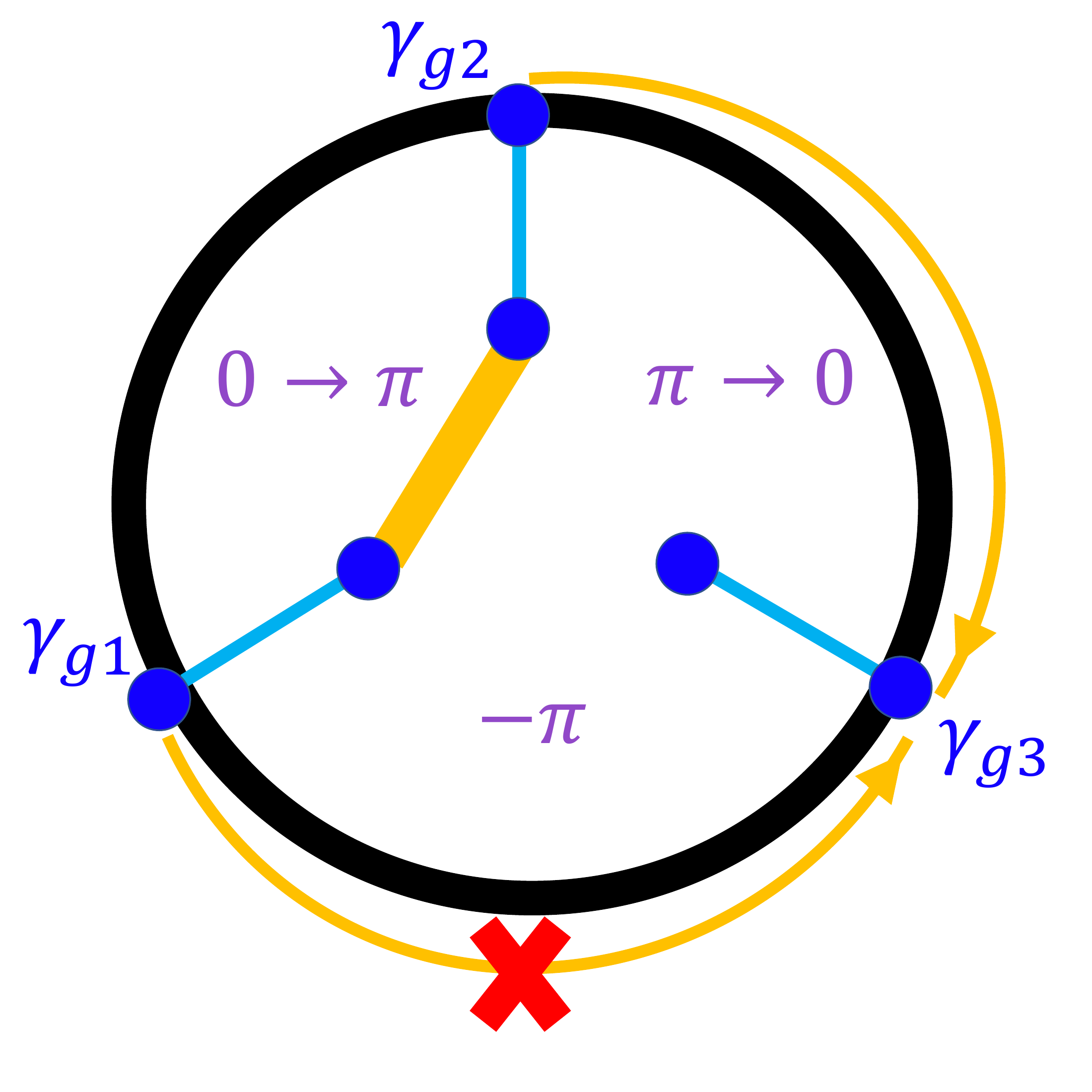}
\end{center}
\vspace{-3mm}
\caption{Depiction of the tri-junction during the first step of the braiding procedure.  The purple numbers show the desired phase difference between nanowires before and after the step.  The goal is to move the Majorana at $\gamma_{g2}$ to $\gamma_{g3}$.  Flux error could instead cause the Majorana at $\gamma_{g1}$ to move to $\gamma_{g3}$}
\label{FR0}
\vspace{3mm}
\end{figure}

Naively, one would think that the device is insensitive to all types of flux error/noise as it is topological. This is not entirely true.  Like all tri-junction devices, it is essential to ensure that Majorana zero modes are braided around each other in the ``right sense.'' This is accomplished by ensuring that there is no stray coupling between Majorana zero modes that are not being exchanged in the particular step of the braiding operation (see Fig.~\ref{FR0}).

%%%%%%%%%%%%%%%%%%%%%%%%%%%%%%%%
%%%%%%%%%%%%%%%%%%%%%%%%%%%%
%%%%%%%%%%%%%%%%%%%%%%	
%%%%%%%%%%%%%%%%%%%%%%%%%%%%%%%%

%%%%%%%%%%%%%%%%%%%%%%%%%%%%%%%%
%%%%%%%%%%%%%%%%%%%%%%%%%%%%
\begin{figure}[t]
\begin{center}
\includegraphics[width=\columnwidth]{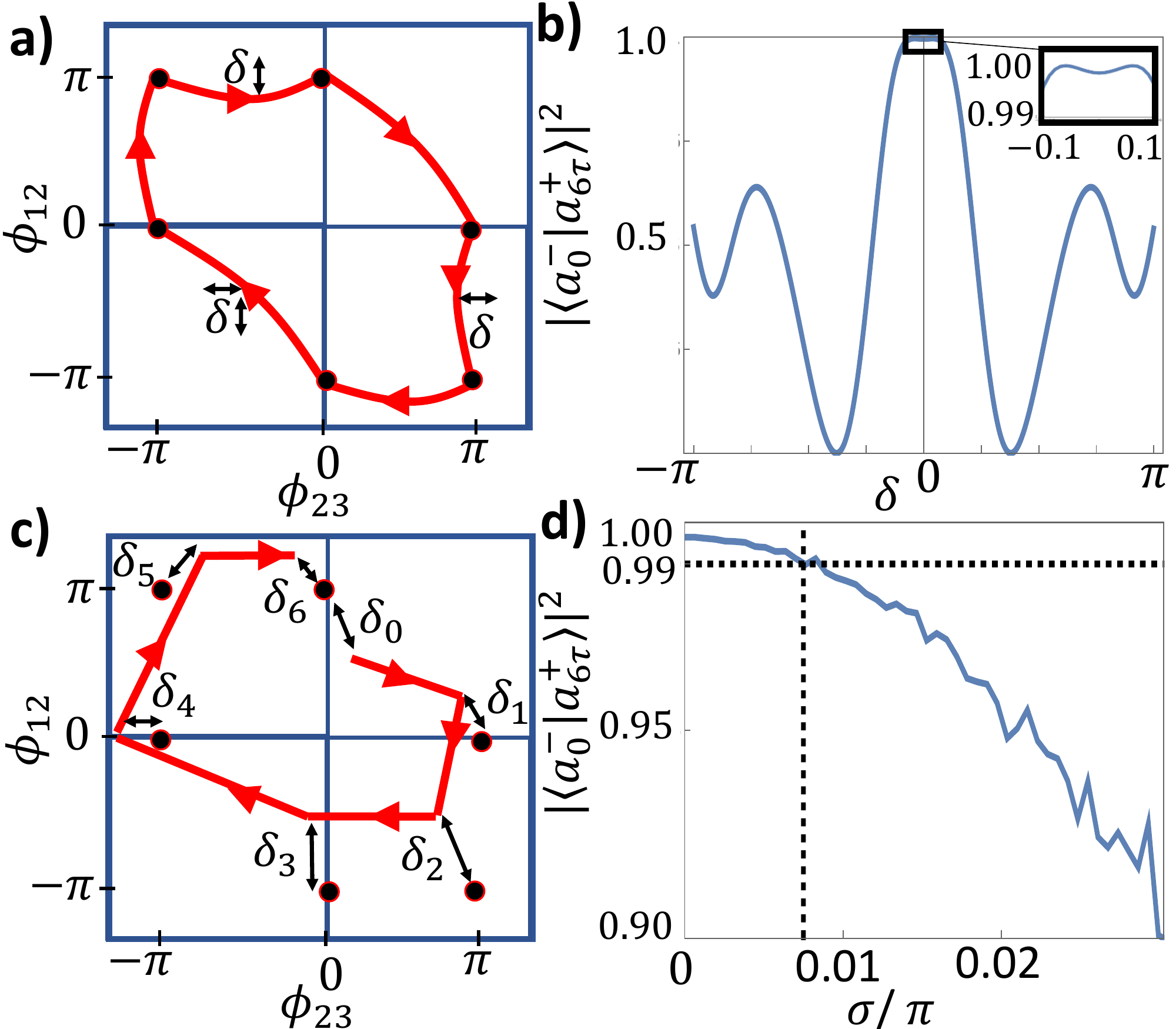}
\end{center}
\vspace{-3mm}
\caption{(a) a path through the space of phase differences that does not go straight from one point to the next.  The parameter $\delta$ characterizes the point along the curve that is furthest from the straight line.  (b)  The fidelity for the path depicted in (a) as a function of the $\delta$ parameter.  The fidelity is flat from about $-\pi/16$ to $\pi/16$ as shown in the offset. (c) a path through the space of phase differences that does not hit the verticies at the end of each step.  Here, $\delta_i$ measures the distance between the point that is hit by the path at the end of step $i$ and the target point.  These parameters $\delta_i$ are generated from a random Gaussian distribution with a standard deviation of $\sigma$.  (d) the fidelity, averaged over 100 trials, for the path depicted in (c) as a function of the standard deviation $\sigma$.}
\label{FR}
\vspace{3mm}
\end{figure}
%%%%%%%%%%%%%%%%%%%%%%	
%%%%%%%%%%%%%%%%%%%%%%%%%%%%%%%%

To demonstrate the effects of flux errors and low frequency flux noise on the device, we consider braiding operations along a deformed trajectory in the $\phi_{12}$-$\phi_{23}$ plane. In type 1 deformations the vertices of the braiding path (where two couplings are completely turned off) are hit but the trajectory connecting these points are not straight lines.  Braiding error occurs if the wrong Majorana is moved during a step in the braiding process due to stray flux turning on the wrong coupling.  As long as the correct couplings dominate, this error is negligible which is the case in type 1 deformations.   Figure~\ref{FR}a depicts an example of type 1 deformation where the paths are deformed from the straight line path $(\phi_{a}(t),\phi_{b}(t))$ in a sinusoidal manner $(\phi_{a}(t)+\delta \sin(t/\tau),\phi_{b}(t)+\delta\sin(t\tau))$ where $\phi_{a}$ and $\phi_b$ are whichever two phase differences that are changing in a particular step, $\tau$ is the duration of the step, and the total phase is kept to zero.    Panel b shows the corresponding fidelity as a function of the magnitude of deformation.  We see that the fidelity plateaus to 1 at relatively large deformations $(|\delta|\approx 0.1)$.     

In type 2 deformations, the verticies of the path are missed.  Type 2 deformations are a greater source of braiding error since, in the vicinity of the verticies, one of the couplings that is being intentionally varied (say $\phi_{23}$) is approaching zero.  Therefore, turning on the third coupling, even a small amount, can cause it to dominate over $\phi_{23}$.  Figure~\ref{FR}c depicts a path in which the target points are missed.  Panel d shows the corresponding average fidelity for random offsets $\delta_i=\sqrt{\delta_{ix}^2+\delta_{iy}^2}$ generated from a two dimensional Gaussian distribution as a function of the standard deviation $\sigma$ of the distribution.  The braiding fidelity stays above 0.99 to about ($\sigma\approx0.007\pi$) which is much smaller than the acceptable deformation in type 1.  However, it is still well above the level of experimental control achievable in the lab.  In the main text, we estimated the duration of the braiding process to be about 600 ns which means we could run the braiding process ten million times and still not see significant flux error from  1/f noise which is about $10^{-12}~\pi$ at 1 Hz~\cite{Bialczak2007}.  Perhaps the largest source of error is the ability of the magnets to set the flux, which is achievable to about 1 part in $10^4$ flux quantum, at which point the average braiding fidelity is approximately 99.999\% of the maximum possible fidelity given perfect flux control.  

\section{Controlling The Phase Differences Via External Flux}
\label{ControllingThePhaseDifferenceViaExternalFlux}

\begin{figure}[t]
\begin{center}
\includegraphics[width=0.4\textwidth]{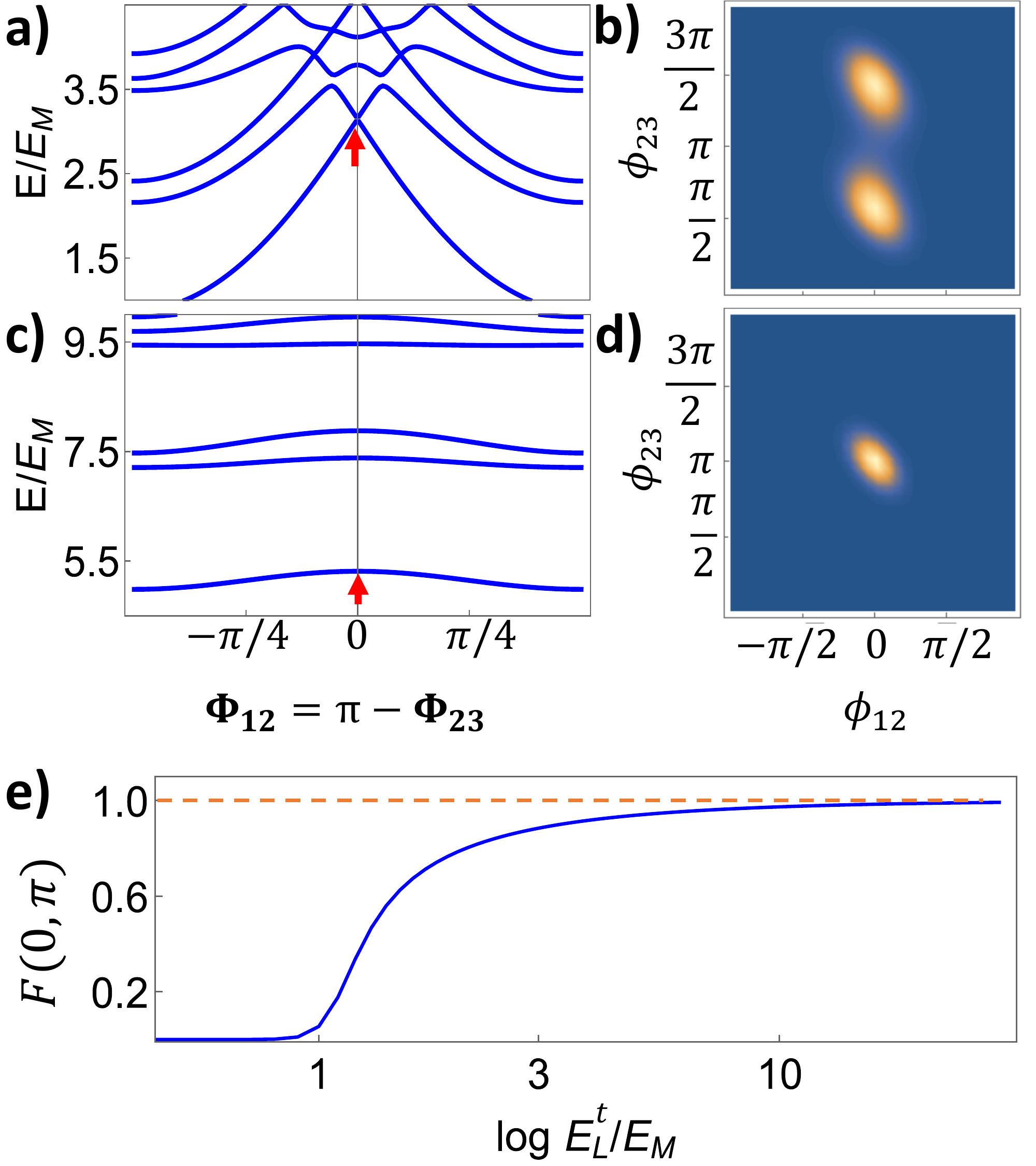}
\end{center}
\vspace{-2mm}

\caption{Controlling the phase difference with the external flux.  Panels (a) and (c) show the energy levels of the braiding ring as a function of the external flux $\Phi_{12}=\pi-\Phi_{23}$ while $\Phi_{31}=-\pi$ for (a) $E_J^t=3.0~E_M$, $E_C^t=0.1~E_M$, and $E_L^t=1.0~E_M$ and (c) $E_J^t=3.0~E_M$, $E_C^t=0.1~E_M$, and $E_L^t=5.0~E_M$.  Panels (b) and (d) show the probability distribution of the ground state as a function of the phase differences $\phi_{12}$ and $\phi_{23}$ corresponding to the red arrow in (a) and (c) respectively.  Panel (e) shows the F-factor which is a measure of the probability of being in the correct state ($\phi_{12}=\Phi_{12}$ and $\phi_{23}=\Phi_{23}$) as a function of the inductive energy.}
\label{F6}
\vspace{-3mm}
\end{figure}

Everything up to this point has assumed that the tri-junction is in the classical limit where the phase difference between superconducting nanowires is completely controlled by the external flux through each loop.  Here we show that the flux control can tolerate a moderate amount of quantum fluctuations.  To see this, we will have to treat the non-Topological part of the Hamiltonian for the tri-junction in a more rigorous manner. 
\begin{equation}
\begin{split}
&H^{t}_{NT}=-\frac{E_C^t}{12} \left( \partial^2_{\phi_{12}}+\partial^2_{\phi_{23}}+(\partial_{\phi_{12}}-\partial^2_{\phi_{23}})^2 \right)
\\
&+E_L^t \left( (\phi_{12}-\Phi_{12})^2+(\phi_{23}-\Phi_{23})^2+(\phi_{12}+\phi_{23}+\Phi_{31})^2 \right)
\\
&-E_J^t \left( \cos(\phi_{12})+\cos(\phi_{23})+\cos(\phi_{12}+\phi_{23}) \right)
\end{split}
\end{equation}
where $E_J^t$ is the Josephson energy for the tri-junction $E_C^t=e^2/2C^t$ is the charging energy, $E_L^t=\hbar^2/4e^2L^t$ is the inductive energy, with $C^t$ and $L^t$ being the capacitance and the inductance of the tri-junction respectively.  The form of the Hamiltonian was derived by applying the flux corral condition $\phi_{12}+\phi_{23}+\phi_{31}=0$ to the classical Lagrangian and finding the canonical momenta before quantizing the Hamiltonian. When the Josephson energy is small and the capacitance is much larger than the inductance then the ring is in the classical regime and the Hamiltonian is well described by the ground state.  However, in general we can solve this Hamiltonian similar to how the non-topological part of the readout loop was solved.  We can change variables and project onto a two dimensional harmonic oscillator basis.  Fig. \ref{F6} was generated in this way keeping the first one hundred ($10 \times 10$) energy levels.  

%%%%%%%%%%%%%%%%%%%%%%%%%%%%%%%%
%%%%%%%%%%%%%%%%%%%%%%%%%%%%
%%%%%%%%%%%%%%%%%%%%%%	
%%%%%%%%%%%%%%%%%%%%%%%%%%%%%%%%

We would like to use the external magnetic flux to control the phase difference between superconductors such that $\phi_{12}=\Phi_{12}$, $\phi_{23}=\Phi_{23}$, and $\phi_{31}=\Phi_{31}$. Since the phase differences follow the condition $\phi_{12}+\phi_{23}+\phi_{31}=0$, we must also set the external flux so that $\Phi_{12}+\Phi_{23}+\Phi_{31}=0$.  In Fig. \ref{F6} panels (a) and (c) we set $\Phi_{31}=-\pi$ and vary $\Phi_{12}$ and $\Phi_{23}$ under the condition $\Phi_{23}=\pi-\Phi_{12}$.  Panels (b) and (d) show the corresponding probability distribution of the ground state over the phases differences $\phi_{12}$ and $\phi_{23}$ at $(\Phi_{12}=0,\Phi_{23}=\pi)$.  For panels (a) and (b) the Josephson energy is the largest of the three energy scales.  We have $E_J^t=3.0~E_M$, $E_L^t=1.0~E_M$, and $E_C^t=0.1~E_M$.  The large Josephson energy causes the phase to delocalize at $(\Phi_{12}=0,\Phi_{23}=\pi)$ where we see that the ground state is not localized to the point  $(\phi_{12}=\Phi_{12}, \phi_{23}=\Phi_{23})$.  In panels (c) and (d), the inductive energy is increased $E_L^t=5.0~E_M$ while all other parameters are kept the same.  There are still some small quantum fluctuations but the probability distribution is now peaked at $(\phi_{12}=\Phi_{12}, \phi_{23}=\Phi_{23})$.  Panel (e) shows the F-factor,
\begin{equation}
F(\Phi_{12},\Phi_{23})=\frac{|\psi_{E_L^t}(\phi_{12}=\Phi_{12},\phi_{23}=\Phi_{23})|^2}{|\psi_{E_L^t\rightarrow\infty}(\phi_{12}=\Phi_{12},\phi_{23}=\Phi_{23})|^2}
\end{equation}
where $\psi_{E_L^t}(\phi_{12},\phi_{23})$ is the wave function for the ground state of the flux part of the braiding ring Hamiltonian with inductive energy $E_L$.  The F-factor is a measure of the probability of the system being at the point $(\phi_{12}=\Phi_{12}, \phi_{23}=\Phi_{23})$.  $F(0,\pi)$ limits to unity as we go to the classical regime (large $E_L$).  However, it stays moderately high even for inductive energies on the order of the Josephson energy.   

\section{Conclusion}
\label{Conclusion}

We have shown that braiding Majorana Zero Modes (MZMs) can be performed using a flux controlled multiterminal $4\pi$ Josephson junction.  Using external magnetic flux, the phase difference between the three arms of a topological junction, and hence the coupling between Majorana Bound States (MBSs) on different arms, can be controlled.  By tuning these couplings we can perform a double braid which flips the sign of odd parity occupation states in the ring.  The state of the topological qubit can be read out via a fluxonium loop embedded with an additional pair of MZMs.  By observing the separation between the ground state and the excited state of the fluxonium, one can test the success of the braiding procedure.  The full device is a complete flux controlled Majorana qubit.  

We thank A. Zarassi for helpful discussions.  Work is supported by NSF PIRE-1743717. S.M.F. is supported by NSF DMR-1743972, ONR, and ARO.

\bibliographystyle{apsrev4-1}
\bibliography{REFERENCES}

\end{document}